\documentstyle[aps,prl]{revtex}

\begin{document}

\title{Stability properties of trapped Bose-Fermi gases mixture}             
\author{X.X.Yi$^{1,2}$, C.P.Sun$^1$}
\address{ $^1$Institute of Theoretical Physics, Academia Sinica, Peking 100080, P.R.China\\
$^2$Institute of Theoretical Physics, Northeast Normal University, Changchun 130024, P.R.China} 
\maketitle
\begin{abstract}
The stability of  Bose-Fermi gases trapped in an isotropic potentials 
at ultracold temperature is
strongly influenced by the interaction between the fermions and the bosons. At
zero temperature, the stability criterion is given in this paper using variation method, 
the results show that 
whether a fermion-boson mixture is stable depends mainly on the interaction between the
fermions and the bosons. For finite temperature, however, the stability is not only 
related to the coupling  constants, but also to the temperature.
The stability conditions for finite temperature 
are also derived and discuss in details in this paper.\\
{\bf PACS number(s):03.75.Fi, 05.30.Fk,05.30.Jp}\\
\end{abstract}
\vskip0.5cm
Since the realization of dilute alkali atomic vapor condensates(Bose-Einstein condensation or BEC)
 in 1995[1], large efforts have been make
to study many-body effects and macroscopic properties of the gases,
which may be more transparently demonstrated in BEC than in other many-body systems. 
For fermionic atomic vapor, however, it is difficult to 
achieve a degenerate gas. Since the  evaporative cooling of a pure fermionic gas is 
ineffective at temperature sufficiently low due to the suppression of $s-$
wave scattering between identical fermions.
As theory
and experiment advanced, a new rich phenomenology has appeared in which 
new conditions arise, which are not 
accessible in other BEC systems. One of the most stunning of these is the recent 
experimental demonstration of a condensate mixture composed of two spin
states of $^{87}Rb$[2]. The realization of two condensates mixture is related 
to the sympathetic cooling mechanism, 
i.e., the exchange of energy due to elastic collisions between atoms of cooled and thermal samples.
Most recently, B.DeMarco and D.S.Jin [3] report their observation of
degenerate Fermi gas using an evaporative cooling strategy. 
Although the strategy uses a two-component
Fermi gas, the mixture of Bose and Fermi gas attracts a lot of attention from 
the viewpoint of both experiment and theoretical study.

The mixed system of Bose and Fermi particles is itself an interesting subject for
investigation. The hydrogen deuterium system has been studied at the early stage of these 
investigations[4], and there is now a lot of literature devoted to the properties of pure 
degenerate trapped atomic Fermi gases[5-9].

In a recent paper, M$\phi$lmer has used a simple mean field models to study
the spatial distribution of a Bose-Fermi gas mixture at $T=0 K$ within Thomas-Fermi approximation.
The results  show that the distributions depend strongly on the relative
sign and magnitude of the boson-boson and boson-fermion scattering lengths.
Here, we shall study the Bose-Fermi gas mixture using a variation method at zero temperature,
this method was first introduced in[10] to study the BEC ground state in a harmonic trap of
a Bose system, and later generalized by H.Shi and W.M.Zheng to study BEC with attractive 
interactions[11]. In addition, we study the stability of the Bose-Fermi gas mixture 
at finite temperature. The results show that there is a region of 
temperature in which the phase 
separation of the mixture happens. And the span of the region depends
on the coupling constants.

To begin, we consider a second-quantized grand canonical Hamiltonian of interacting 
Bose and Fermi gases
\begin{eqnarray}
H&=&H_b+H_f+V_{bf},\nonumber\\
H_b&=&\int dr\phi^+(r)(\frac{p^2}{2m_b}-\mu_b+\frac 1 2 m_b\omega_br^2)\phi(r)+\frac {g_{bb}} {2} 
\int\int dr dr^{'}\phi^+(r)\phi^+(r^{'})\phi(r^{'})\phi(r),\nonumber\\
H_f&=&\int dr\psi^+(r)(\frac{p^2}{2m_f}-\mu_f+\frac 1 2 m_f\omega_fr^2)\psi(r),\nonumber\\
V_{bf}&=&g_{bf}\int dr dr^{'}\phi^+(r)\psi^+(r^{'})\delta(r-r^{'})\psi(r^{'})\phi(r),
\end{eqnarray}
where $\phi(r)$ and $\psi(r)$ denote boson and fermion field operators with
masses $m_b$ and $m_f$, respectively.
For weakly interacting dilute gases, the interactions between the bosonic atoms are 
modeled by $\delta$ potentials and the interactions among
the fermionic atoms are neglected, since the interactions between atoms at very low temperature is 
suppressed for polarized systems.
$g_{bb}$ and $g_{ba}$ stand for boson-boson and boson-fermion coupling constant, respectively.
$$g_{bb}=\frac{4\pi\hbar^2}{m_b}a_{bb}, g_{bf}=\frac{2\pi\hbar^2}{m_{bf}}a_{bf},$$
$a_{bb}$($a_{bf}$)  are $s$-wave scattering length between
boson and boson (boson and fermion), and $m_{bf}$ is a reduced mass of the boson
and the fermion. The chemical potentials $\mu_b$ and $\mu_f$ are determined through the conditions
\begin{equation}
N_b=\langle\int dr \phi^+(r)\phi(r)\rangle, N_f=\langle\int dr\psi^+(r)\psi(r)\rangle.
\end{equation}
At $T=0$, self-consistent mean field theory, assuming
that all $N$ bosonic particles in a gas populated the same state denoted by 
single particle wave function
$\Phi(r)$, lead to a nonlinear Schr\"odinger equation (or the Gross-Pitaevskii equation) for
$\Phi(r)=\langle\phi(r)\rangle$
\begin{equation}
[-\frac{\hbar^2}{2m_b}\bigtriangledown^2 +\frac 1 2 m_b\omega_b^2r^2+g_{bb}n_b(r)]\Phi(r)=E_b\Phi(r),
\end{equation}
we here omit quantities $g_{bf}n_f(r)$, which is smaller than $g_{bb}n_b(r)$ in 
the case of $N_b>>N_f$.
In order to get a degenerate fermionic gas, 
the boson particles appear in the system only as a coolant, so the number of 
bosons is always much larger than the number of fermions. In the same approximation, the
fermionic wave function is given by a Slater determinant
\begin{equation}
\Psi(r_1,r_2,...,r_{N_f})=\frac{1}{\sqrt{N_f!}}
\left[
\begin{array}{cccc}
\Psi_1(r_1) & \Psi_1(r_2) & \cdot \cdot \cdot &\Psi_1(r_{N_f})\\
\Psi_2(r_1) & \Psi_2(r_2) & \cdot \cdot \cdot &\Psi_2(r_{N_f})\\
\cdot   &  \cdot   &  \cdot \cdot \cdot &  \cdot \\
\cdot  &  \cdot  &  \cdot \cdot \cdot &  \cdot \\
\cdot &  \cdot  &  \cdot \cdot \cdot  &  \cdot  \\
\Psi_{N_f}(r_1) & \Psi_{N_f}(r_2) &\cdot \cdot \cdot &\Psi_{N_f}(r_{N_f})
\end{array}
\right ],
\end{equation}
where $\Psi_i(r)$ is the single particle states determined by Hartree-Fock self-consistent equation
\begin{equation}
[-\frac{\hbar^2}{2m_f}\bigtriangledown^2 +\frac 1 2 m_f\omega_f^2r^2+g_{bf}n_b(r)]\Psi_i(r)=
E_i\Psi_i(r).
\end{equation}
The  density of the fermions is given by
\begin{equation}
n_f(r)=|\Psi(r)|^2.
\end{equation}
In the semiclassical (Thomas-Fermi) approximation, the particle are
assigned classical position and momenta, but the effects of quantum statistics are
taken into account. Under this approximation, the Eqs.(3) and (5) for the boson and fermion
wave function are equivalent to[13,14]
\begin{eqnarray}
\frac 1 2 m_b\omega_b^2r^2+g_{bb}n_b(r)&=&\mu_b,\nonumber\\
\frac{\hbar^2}{2m_f}[6\pi^2n_f(r)]^{\frac 2 3 }+\frac 1 2 m_f\omega_f^2r^2+g_{bf}n_b(r)&=&e_F.
\end{eqnarray}
The main conclusion of this equations is discussed in Ref.[13].
We obtain $n_b(r)=\frac{1}{g_{bb}}(\mu_b-\frac 1 2 m_b\omega_b^2 r^2)$ from
the first line of Eqs(7). Substituting $n_b(r)$ into the second line of Eqs(7), we
yield
\begin{equation}
\frac{\hbar^2}{2m_f}[6\pi^2n_f(r)]^{\frac 2 3}+\frac 1 2 m_f\omega^2_fr^2+
\frac{g_{bf}}{g_{bb}}(\mu_b-\frac 1 2 m_b\omega_b^2r^2)=e_F,
\end{equation}
this equation shows that the fermions experience a potential minimum in the center of the trap
if $g_{bf}/g_{bb}<m_f\omega_f^2/m_b\omega_b^2,$ in this case the entire
distribution behaves like a fermionic core within the Bose condensate. The fermion density is
a constant throughout the Bose condensate if $g_{bf}/g_{bb}=m_f\omega_f^2/m_b\omega_b^2.$ 
Whereas the fermions are repelled from the center of the trap and 
localized near the edge of the Bose condensate if $g_{bf}/g_{bb}>m_f\omega_f^2/m_b\omega_b^2,$
i.e. a phase separation occurs in this system. 
We would like to note that the distribution of BEC remains unchanged in the above discussions,
since we assume $N_b>>N_f$.
To drive Eq.(8), we assume that the Thomas-Fermi approximation(TFA) is valid. The coupling constant
$g_{bb}$ and $g_{bf}$ may take any value as long as the TFA is available, and the phase separation 
depend mainly on ratio $g_{bf}/g_{bb}$. In what follows 
we discuss the separation of the bosonic and fermionic parts from the other aspect for zero
temperature by using variation method, the results are indeed different from those under TFA.
We note the solution of Eq.(5) requires prior knowledge of the boson density profile $n_b=|\Phi(r)|^2$. 
To obtain the density profile, we have to solve the Gross-Pitaevskii equation (3). There are a 
large number of literatures devoted to solve the Gross-Pitaevskii equation[12], we here
use a variation method[11] to solve the problem.
For a isotropic trapping potential, we may assume the trial wave function for
$\Phi(r)$ in Eq.(3) to be 
\begin{equation}
\Phi(r)=\sqrt{N_b}\omega^{\frac 3 4 }(\frac{m_b}{\pi\hbar})^{\frac 3 4 }
e^{-m_b\omega r^2/2\hbar},
\end{equation}
where $\omega$ is the effective frequency and  is taken as a variational parameter. Substituting
Eq.(9) into Eq.(3), we obtained the ground-state energy
\begin{equation}
E_b[\Phi]=E_b(\omega)=\frac 3 4 N_b\hbar\omega+\frac 3 4 N_b\hbar
\frac{\omega_b^2}{\omega}+g_{bb}N_b^2(\frac{\omega m_b}{2\pi\hbar})^{\frac 3 2}.
\end{equation}
If $E_b(\omega)$ is plotted as a function of $\omega$ , one sees that a stable local 
minimum exists only up to a certain maximum number of atoms for $g_{bb}<0$[11]. The critical point
occurs where
\begin{equation}
\frac{\partial E_b(\omega)}{\partial \omega}|_{(\omega=\omega_c,N_b=N_{bc})}=0, \mbox{\ \ and\ \ }
\frac{\partial^2 E_b(\omega)}{\partial \omega^2}|_{\omega=\omega_c,N_b=N_{bc}}>0.
\end{equation}
Here, $\omega_c$ stands for the variational parameter that minimizes the ground state energy. 
Using equation (10), for $g_{bb}<0$ the critical number of bosons is given by
\begin{equation}
N_b^c<2\hbar\omega_b^2(\omega_c)^{-\frac 5 2}\frac{1}{|g_{bb}|}(\frac{2\pi\hbar}{m_b})^{\frac 3 2}.
\end{equation}
where $\omega_c$ satisfies
\begin{equation}
\hbar\omega_c^2-\hbar\omega_b^2+2g_{bb}N_b(\frac{m_b}{2\pi\hbar})^{\frac 3 2 }\omega_c^{\frac 5 2 }
=0.
\end{equation}
Parameter $\omega_b\sim 166Hz$ relevant to the experiment gives  $N_b^c\sim 1400$, which
is in good agreement with the experiment[1,11].
The solution $\omega_c$ of Eq.(13) against $g_{bb}$ is plotted in Fig.1, which shows that 
as $|g_{bb}|$ increases, the variation parameter  $\omega_c$ decreases, and it has 
a maximum equal to $\omega_b$ at $g_{bb}=0$. We will use this solution to study the stability
of the mixture at zero temperature below.

We may determine the ground-sate energy functional of the fermions provided $n_b(r)$ is known. 
In terms of the fermion distribution $n_f(r)$, the energy functional $E_f$ of the
fermions is given by[15]
\begin{equation}
E_f=E_f[n_f(r)]=\int \frac{d^3r}{6\pi^2}\frac{\hbar^2}{2m_f}[6\pi^2n_f(r)]^{5/3}+
\frac 1 2 \int d^3r m_f\omega_f^2r^2n_f(r)+\int d^3r g_{bf}n_f(r)n_b(r),
\end{equation}
since the interaction between the bosons and the fermions is rather week,
we may consider a Gaussian function as a trial fermions' distribution
\begin{equation}
n_f(r)=N_f\Omega^{3/2}(\frac{m_f}{\pi\hbar})^{3/2}e^{-\frac{m_f\Omega(r-r_f)^2}{\hbar}}.
\end{equation}
Here $r_f$ and $\Omega$ is treated as
variation parameters.
Substituting this wave functions into Eq.(14), one obtains
\begin{equation}
E_f=E_f(\Omega,r_f,N_f,\omega_c)=P+\frac 1 2 m_f\omega_f^2r_f^2N_f+
hN_bN_f(\pi G)^{\frac 3 2 }e^{-Gr_f^2}
\end{equation}
with
$$P=P(\Omega,N_f)=(\frac 3 5)^{\frac 3 2 }\frac{1}{\pi}(6\pi^2)^{\frac 2 3 }\hbar
\Omega N_f^{\frac 5 3}+\frac{3\hbar \omega_f^2}{4\Omega}N_f$$
$$G=G(\Omega,\omega_c)=\frac{m_fm_b\omega_c\Omega}{\hbar(m_f\Omega+m_b\omega_c)}.$$
As known, a physical state corresponds to a stable or metastable point of the 
energy functional. If a separation of the fermion and boson component occurs, then $r_{fc}$
that minimizes the energy $E_f$ takes a positive nonzero value. In other words, there
are no separations between the two components when energy $E_f$ exhibits a minimal value at
$r_f=0$. When distribution function is restricted to the form of the trial function (15)
we may write the conditions of a minimal energy in terms of derivatives of the 
energy with respect to the adjustable variation parameters of the trial function.
We show them as follows
\begin{eqnarray}
\frac{\partial E_f}{\partial \Omega}|_{\Omega=\Omega_c}=0,
\frac{\partial E_f}{\partial r_f}|_{r_f=r_{fc}}=0,\nonumber\\
\frac{\partial^2E_f}{\partial \Omega^2}
\frac{\partial^2 E_f}{
\partial r_f^2}-(\frac{\partial^2 E_f}{\partial\Omega\partial r_f})^2>0,
\end{eqnarray}
the stationary conditions (there are no separation between the boson and fermion
) are
\begin{eqnarray}
Y&=&Y(N_b,\omega_c,g_{bf})=[\hbar\frac{\omega_f^2}{\Omega_c^3}+g_{bf}
N_b\pi^{\frac 3 2 }\sqrt{G(\Omega_c,\omega_c)}\frac{\partial^2 G(\Omega_c,\omega_c)}
{\partial \Omega^2}+\frac{g_{bf}N_b\pi^{3/2}}{2G(\Omega_c,\omega_c)}
(\frac{\partial G(\Omega_c,\omega_c)}{\partial\Omega})^2]\nonumber\\
&\times&[m_f\omega_f^2-2g_{bf}N_b\pi^{3/2}
G^{5/2}(\Omega_c,\omega_c)]>0,
\end{eqnarray}
where $\Omega_c$ determined by $\partial E_f/\partial \Omega|_{\Omega_c}=0$ 
satisfies the following equation
\begin{equation}
\frac{2}{3\pi}(6\pi^2)^{\frac 2 3 }(\frac 3 5 )^{\frac 3 2 }\hbar N_f^{\frac 2 3 }
-\frac 1 2 \hbar (\frac{\omega_f}{\Omega_c})^2+ g_{bf}N_b\pi^{\frac 3 2 }
\sqrt{G(\Omega_c,\omega_c)}\frac{\partial G(\Omega_c,\omega_c)}{\partial \Omega}=0,
\end{equation}
the solution of Eq.(19) as a function of $g_{bf}$  is shown in Fig.2, a magnification part of the curve near
$g_{bf}=0$ (but $g_{bf}>0$) is give in the inset. This
curve indicates that the fermions prefer to occupy the trap centre
for $g_{bf}<0$, and the larger the coupling constant 
$|g_{bf}|(g_{bf}<0)$, the sharper the distribution 
of the fermions. 
However, as we show below, the fermions and the bosons can not always coexist even if $g_{bf}<0$.
$\frac{\partial E_f}{\partial r_f}|_{r_f=r_{fc}}=0$ has two solutions, one solution
is $r_{fc}=0$ and the another is 
\begin{equation}
r_{fc}=\sqrt{\frac{1}{G}ln(2g_{bf}N_b\pi^{\frac 3 2 }G^{\frac 5 2}(\Omega_c,\omega_c))-
ln(m_f\omega_f^2)}.
\end{equation}
 For
$g_{bf}<0$ or $$0<g_{bf}<\frac{m_f\omega_f^2}{2N_b\pi^{\frac 3 2}G^{\frac 5 2 }(\Omega_c,\omega_c)}$$
i.e. the interaction between fermion and the boson is attractive or weekly repulsive,
the solution  $r_{fc}=0$ holds, which indicate that there is not separation between the
fermions and the bosons. For
$g_{bf}>\frac{m_f\omega_f^2}{2N_b\pi^{\frac 3 2}G^{\frac 5 2 }(\Omega_c,\omega_c)}$
the fermions experience a effective potential minimum at 
$r_{fc}=\sqrt{\frac{1}{G}ln(2g_{bf}N_b\pi^{\frac 3 2 }G^{\frac 5 2}(\Omega_c,\omega_c))-
ln(m_f\omega_f^2)}>0, $
the Bose condensate is surrounded by a shell of fermions in this case.
$Y(N_b,g_{bf},\omega_c)$ 
as functions of the coupling constant $g_{bf}$ are shown in figure 3. We see that
whether the boson-fermion mixture is stable depends not only
on the coupling constant $g_{bf}$ and $g_{bb}$(through $\omega_c)$, but also on $N_b$ and 
$N_f$(through $\Omega_c$), i.e., the 
stability of the mixture system depends on the 
number of both boson and fermion system. For example, in Fig.3-a we show 
$Y$ given by Eq.(8) as a function of the coupling constant $g_{bf}$ for fixed
$N_f=100, N_b=10000$, while Fig.3-b is for the same parameters 
as in Fig.3-a except for $N_b=1000$, it is obvious that the region of $g_{bf}$ in which the system
has no phase separation has been broadened with $N_b$ decreases (for fixed $N_f$). 
The inset present the dependence of $Y$ on $g_{bf}$ at a larger scale of $g_{bf}.$
It is interesting to compare the above mentioned 
results with those obtained by treating the fermions in the Thomas-Fermi approximation,
this is done in Ref.[13,14], and we note that the semiclassical description gives a qualitatively 
correct description and it reliably predicts the phase separation.

Now we tune our attention to discuss the above problem at finite temperature. First of all, we 
consider the homogenous case, for
the boson and fermion system, thermodynamical properties are trivial if there are not 
interaction between them. But in this case the sympathetic cooling scheme 
does not take any effect and the degenerate fermions in a trapped potential
have not been achieved. The thermodynamical properties may be changed when
the interaction between the fermions and bosions is turn on, then a new phenomenon,
the phase separation, may occur in this system. for a homogeneous fermion and boson mixture
system, the Helmholts free energy can be written as[16]
\begin{equation}
\beta F=-\frac{V}{\lambda_2^3}f_{\frac 5 2 }(z_f)+\frac 1 2 g_{ff}\rho_fN_f\lambda_f^2+ln(1-z_b)-
\frac{V}{\lambda_b^3}g_{\frac 5 2}(z_b)+2g_{bb}\rho_bN_b\lambda_b^2+
g_{bf}(\lambda_b^2+\lambda_f^2)N_fN_b/V,
\end{equation}
where index $f$ refers to the fermionic component, whereas index $b$ stands for the 
bosonic one, $N_i$ is the 
number of particles in component $i$, $\lambda_i$ denotes the thermal wave 
length of component $i$, $f_n(z)$ and $g_n(z)$ represent the Fermi and Bose
integral, respectively.The equation (21) is based on the pseudopotential form
of the atom-atom interaction, and may be assumed accurate when the system is dilute.
i.e. $\rho_i g_{ii}^3<<1$ and $g_{ii}/\lambda_i<<1$, where $\rho_i$ is the density of the component $i$.
This condition is well satisfies for the samples of alkali atoms in experiments to date[1,12,17,18].

 From eq.(21) we obtain the chemical potential for each
component straightforwardly,
\begin{eqnarray}
\beta\mu_b&=&\beta\mu_b^0+4g_{bb}\rho_{b}\lambda_b^2+g_{bf}(\lambda_b^2+\lambda_f^2)
N_f/V,\nonumber\\
\beta\mu_f&=&\beta\mu_f^0+g_{ff}\rho_f\lambda_f^2+g_{bf}(\lambda_b^2+\lambda_f^2)
N_b/V,
\end{eqnarray}
where $\mu_i^0$ are the chemical potentials of ideal gas. There
are three terms in each chemical potential, the 
second term comes from the interaction within the component and the third 
term is from the interaction  between the fermion and boson component.
As known, an homogenous binary mixture is stable only when the symmetric matrix
$\hat{\mu}$ given by
\begin{equation}
\hat{\mu}=
\left[
\begin{array}{ll}
\frac{\partial\mu_b}{\partial\rho_b} & \frac{\partial\mu_b}{\partial\rho_f}\\
\frac{\partial\mu_f}{\partial\rho_b} & \frac{\partial\mu_f}{\partial\rho_f}
\end{array}
\right ]
\end{equation}
is non-negatively definite, in other words, all 
eigenvalues of matrix $\hat{\mu}$ given in Eq.(23)  are non-negative. Mathematically,
for homogeneous fermion and boson mixture the stability conditions are
\begin{equation}
\frac{\partial \mu_b}{\partial\rho_b}\geq 0,\frac{\partial \mu_f}{\partial\rho_f}\geq0,
\end{equation}
and 
\begin{equation}
det\left [
\begin{array}{ll}
\frac{\partial\mu_b}{\partial\rho_b} & \frac{\partial\mu_b}{\partial\rho_f}\\
\frac{\partial\mu_f}{\partial\rho_b} & \frac{\partial\mu_f}{\partial\rho_f}
\end{array}
\right ] \geq 0.
\end{equation}
For ideal gas, we have $\rho_b=\frac{1}{\lambda_b^3}g_{\frac 3 2 }(z_b),$
$\rho_f=\frac{1}{\lambda_f^3}f_{\frac 32 }(z_f),$
this leads to
\begin{equation}
\beta\frac{\partial\mu_f^0}{\partial\rho_f}=\frac{\lambda_f^3}{f_{\frac 1 2 }(z_f)},\,
\beta\frac{\partial\mu_b^0}{\partial\rho_b}=\frac{\lambda_b^3}{g_{\frac 1 2}(z_b)},
\end{equation}
It follows from eqs (24) and (25) that
\begin{eqnarray}
4g_{bb}\lambda_b^2+\frac{\lambda_b^3}{g_{\frac 1 2 }(z_b)}\geq 0,\\
g_{ff}\lambda_f^2+\frac{\lambda_f^3}{f_{\frac 1 2 }(z_f)}\geq 0,\\
\mbox{and}\nonumber\\
Z(T,g_{bf},g_{ff},g_{bb})=Z=(4g_{bb}\lambda_b^2+\frac{\lambda_b^3}{g_{\frac 1 2 }(z_b)})(g_{ff}
\lambda_f^2+
\frac{\lambda_f^3}{f_{\frac 1 2 (z_f)}})-g_{bf}^2(\lambda_b^2+\lambda_f^2)^2
\geq 0.
\end{eqnarray}
It is well known that a homogeneous imperfect gas with attractive interaction is not stable.
The fermions in this kind of gas could form BCS state, which consists two fermionic 
particles interacting with each other but not with the other fermions from the Fermi gas, whereas
bosons with attractive interaction could  collapse into liquid. Hence, we here discuss the system
with repulsive interactions. It is obvious that the stability condition (27) and (28) 
hold always for 
$g_{bb}>0$, $g_{ff}>0$. We would like to point out that the stability conditions (27-29) 
do not involve 
the densities of the both components. At first sight, this seems to be confusion, in fact, there
is no contradiction. One can demonstrate that at low density the Helmholtz free energy of the
bogoliubov gas reduce to a quadratic form in $N_b$ and $N_f$.
To have a minimum, this form should be positive definite, i.e.,
$\mbox{det}||\frac{\partial^2 F}{\partial N_b\partial N_f}||\geq0.$ Therefore, the 
corresponding stability criterion involves
only density-independent constants in the order of
approximation used. This criterion is similar to the
stability conditions for two-component Bose-Einstein condensate in a trapped
untracold gas[19-25]. When $T\rightarrow \infty, \lambda_i\rightarrow 0$,
hence $Z\sim \frac{1}{\rho_b\rho_f}$. Thus at high temperature,
the homogeneous binary gas mixture is always stable and no phase
separation occur. In the case considered here, Fermi
temperature $T_F=\frac{h^2}{2mk_B}(\frac{3N_f}{8\pi V})^{\frac 2 3}$ 
is much lower than BEC temperature $T_c=\frac{h^2}{2\pi m K_B}(\frac{N_b}{2.612V})^{\frac 2 3},$
 i.e.,as temperature decreases, it first passes the BEC transition point $T_c$.
When $T\rightarrow T_c$, $g_{\frac 1 2 }(1)\rightarrow \infty$, so
$$Z(T,g_{bf},g_{bb},g_{ff})\sim 4g_{bb}\lambda_b^2(g_{ff}\lambda_{f}^2+\frac{\lambda_f^3}{f_{\frac 1 2 }(z_f)})
-g_{bf}^2(\lambda_b^2+\lambda_f^2)^2.
$$
In particular, when $T<<T_F$,i.e., the temperature is much smaller than
the Fermi temperature of the fermion system,  the stability condition becomes
(setting $m_f=m_b$)
\begin{equation}
g_{bb}g_{ff}-g_{bf}^2\geq 0,
\end{equation}
which does not depend on temperature and coincides 
with the stability conditions of two-component BEC[19,25]. 
Although it
is difficult to reach this region of very low
temperature, yet it attracts much more attention. Because both superfluidity
and shell effects are expected to occur at temperature much smaller than the Fermi temperature[6,26].
$Z$ given by Eq.(29) as a function of the temperature is shown in Fig.4, we see that
the system is always stable when $T\rightarrow 0$ and $T\rightarrow \infty$, and the system is
unstable for $T_{c1}<T<T_{c2}$, where $T_{c1}$ and $T_{c2}$ are roots of 
$Z(T,g_{bb},g_{bf},g_{ff})=0$. In particular, $T_{c1}$ and $T_{c2}$ depend on $g_{bf}$
 $g_{ff}$ and $g_{bb}$. As $g_{bf}$ decreases (for fixed $g_{bb}$ and $g_{ff}$), $T_{c1}$
tends to $T_{c2}$ (in Fig.4 going from dotted line to solid line). 
The critical temperature $T_{c1}$ and $T_{c2}$ characterize the onset of the phase separation,
which is quite different from the Bose-Einstein condensation and the degenerate fermions. 
The critical temperature of BEC and of the onset of degenerate fermionic gas depend mainly on 
the density of the system $N_i/V( i=b,f)$. Especially, the BEC and the degenerate
fermionic gas may happen even if $g_{bf}=0$. For the phase separation, however, nothing will happen if 
$g_{bf}=0$. For a fixed temperature and the coupling constant $g_{bf}$, $Z$ vs. $g_{bb}$ and $g_{ff}$
 is 
shown in Fig.5, which represents the dependence of the stability on the 
interaction strength inside each component.

Until now, we considered only a homogeneous Fermi-Bose gas mixture at finite temperature. 
In reality, however, experiments 
with ultracold atoms are performed by trapping and cooling in an external potential
that can be generally modeled by an isotropic harmonic oscillator $V(r)=\frac m 2\omega_t^2 r^2$,
where $\omega_t$ is the trapping frequency. An
exact criterion for the stability of an inhomogeneous Bose-fermi mixture should involve calculating 
the Helmholts free energy as a function at all
eigenstates of the trapping potential. Fortunately, in the system
considered here it is a good approximation to take use of the 
local-density approximation, which treats the system as being locally homogeneous. This requires that
the level spacing $\hbar\omega_t$ of the trapping potential is much smaller
than the Fermi energy. Of course, the local density approximation always breaks down at the 
edge of the gas cloud where the density vanishes and the effective Fermi energy becomes zero.
In this approximation, the stability conditions can still be
calculated by means of the equations derived above, with the understanding that now the 
effective chemical potentials are spatially dependent through
$$\mu_b=\mu_b^0-\frac 1 2 m_b\omega_b^2r^2,\, \mu_f=\mu_f^0-\frac 1 2 m_f\omega_f^2r^2.$$
Thus a local stability condition is the same as given in Eq.(29) but replacing $z_i(i=1,2)$
by
$$
\tilde{z}_1=z_1e^{-\frac{\beta}{2}m_b\omega_b^2r^2},\mbox{\ \ and\ \ }
\tilde{z}_2=z_2e^{-\frac{\beta}{2}m_f\omega_f^2 r^2}.$$
As shown in inset of Fig.4, 
the region of temperature in which the system is unstable decrease for the case
of $r\neq 0$. 
As compared with the case without trapped potential, the total energy of the system increases 
for it in a trap. Alternatively, within the TFA, the chemical potentials decrease in this process.
So this effect is equal to be that the particle number of the system has a loss.
In this sense, the system is more stable than before.

In summary, we considered a dilute Bose-Fermi gas mixture in an isotropic trap. The atom can interact via
s-wave scattering except within the fermions. These interactions strongly affect the 
stability of the system at zero and finite temperature. In addition, the stability conditions 
depend on the ratio rate $N_b/N_f$, the larger the ratio rate, the smaller the 
region of stability.
For finite temperature, however, the stability conditions depends not only on the 
interactions,but also on the temperature. The region $T_{c1}\leq T\leq T_{c2}$ in which 
the system is unstable depend on the strength of the interaction between and within the 
bosons and the fermions. For an anisotropic trap, the stability conditions 
remain unchanged, whereas somewhat would be changed for zero temperature compared with
the case of isotropic trap. To study the effects, we should introduced the another variation parameter 
in Eqs (9) and (5) 
to characterize the BEC and the fermions in this trap. Consequently, the stability condition
(18) for zero temperature changes and the phase separation could different for different orientation. 
These need further investigations.
\\
{\bf \large ACKNOWLEDGEMENT:}\\
We thank Dr. Li You for his stimulating and helpful discussions.\\

{\bf Figure captions:}\\
Fig. 1:The parameter $\omega_c$ which minimizes the energy functional 
versus the coupling constant $g_{bb}$. The trapped frequency 
$\omega_b=166 Hz$ and the  number of the bosonic atom $N_b=1000$.\\
\ \ \\
Fig. 2:The parameter $\Omega_c$ which minimizes the energy functional 
as a function of the coupling constant $g_{bf}$. 
The parameters chosen are $g_{bb}=0.05$ in units of $\hbar\omega_f a^3$ 
($a=\sqrt{\hbar/\omega_bm_b}$)
and all the coupling constants are chosen in this units hence forth, $N_f=100$, $\omega_f=166Hz$.
Scatter and solid line correspond to different number of bosonic atom,
as specified in the figure. The inset presents the enlarged part of the curve near $g_{bf}=0(>0)$.\\
\hspace{0.3cm}
\ \ \\
Fig.3:Plot of $Y$ given by Eq.(18) as a function of coupling constant $g_{bf}$.
The parameter chosen are a:$N_f=100, N_b=1000.$ b:$N_f=100, N_b=10000.$ The curve for a larger 
scale of $g_{bf}$ is presented as an inset in the figure.\\
\hspace{0.3cm}
\ \ \\
Fig.4:Plot of $Z$ given by Eq.(29) as a function of temperature $T$. The parameters chosen are 
$N_b=1000, N_f=10000, g_{bb}=0.05, g_{ff}=0.01$. Dashed-dotted line:$g_{bf}=0.3$, dotted line
$g_{bf}=0.02$, solid line $g_{bf}=0.01$. The dotted line in the inset is the same as the dotted
line in the figure, while the solid line in the inset is for the gases in a trap with trapped 
frequency $166Hz$.\\
\hspace{0.3cm}
\ \ \\
Fig.5:Plot of $Z$ as a function of $g_{ff}$ and $g_{bb}$. The parameters chosen are
temperature $T=0.1 T_F$, $g_{bf}=0.2$. 

\end{document}